\magnification \magstep1
\raggedbottom
\openup 1\jot
\voffset6truemm
\def\cstok#1{\leavevmode\thinspace\hbox{\vrule\vtop{\vbox{\hrule\kern1pt
\hbox{\vphantom{\tt/}\thinspace{\tt#1}\thinspace}}
\kern1pt\hrule}\vrule}\thinspace}
\headline={\ifnum\pageno=1\hfill\else
\hfill {\it Axial Gauge in Euclidean Quantum Gravity}
\hfill \fi}
\centerline {\bf AXIAL GAUGE IN EUCLIDEAN QUANTUM GRAVITY}
\vskip 1cm
\centerline {\it Ivan G. Avramidi,$^{1}$ Giampiero Esposito$^{2,3}$
and Alexander Yu. Kamenshchik$^{4}$}
\vskip 1cm
\noindent
${ }^{1}$Department of Mathematics, University of Greifswald,
Jahnstr. 15a, 17487 Greifswald, Germany
\vskip 0.3cm
\noindent
${ }^{2}$Istituto Nazionale di Fisica Nucleare, Sezione di
Napoli, Mostra d'Oltremare Padiglione 20, 80125 Napoli, Italy
\vskip 0.3cm
\noindent
${ }^{3}$Dipartimento di Scienze Fisiche, Mostra d'Oltremare
Padiglione 19, 80125 Napoli, Italy
\vskip 0.3cm
\noindent
${ }^{4}$Nuclear Safety Institute, Russian Academy of Sciences,
52 Bolshaya Tulskaya, Moscow 113191, Russia
\vskip 1cm
\noindent
{\bf Abstract.} The axial gauge is applied to the analysis of
Euclidean quantum gravity on manifolds with boundary. A set
of boundary conditions which are completely invariant under
infinitesimal diffeomorphisms require that spatial components
of metric perturbations should vanish at the boundary, jointly
with all components of the ghost one-form and of the 
gauge-averaging functional. If the latter is taken to be of
the axial type, all components of metric perturbations obey
Dirichlet conditions, and all ghost modes are forced to vanish
identically. The one-loop divergence coincides with the
contribution resulting from three-dimensional 
transverse-traceless perturbations.
\vskip 100cm
The recent attempts to obtain a correct formulation of
Euclidean quantum gravity on manifolds with boundary [1]
have shed new light on the quantization programme in
covariant or non-covariant gauges. The consideration of
boundaries is suggested by the problems of quantum field
theory (e.g. Casimir effect, van der Waals forces) and
by the attempts to define the quantum state of the universe [1].
Whenever boundaries occur, one would like to ensure that the
whole set of boundary conditions are invariant under 
infinitesimal gauge transformations, since the underlying
classical theory has this property, and perturbative quantum
theory may be viewed as a theory of small disturbances 
around some background-field configurations. In the case
of the gravitational field, which is the object of our
investigation, the whole set of metric perturbations $h_{ab}$
are subject to the infinitesimal gauge transformations
${ }^{\varphi}h_{ab} \equiv h_{ab}+ \nabla_{(a} \; 
\varphi_{b)}$, where $\nabla$ is the Levi-Civita connection
of the background four-geometry with metric $g$, and
$\varphi_{\nu}dx^{\nu}$ is the ghost one-form. For problems 
with boundaries, one thus finds
$$
{ }^{\varphi}h_{ij}=h_{ij}+\varphi_{(i \mid j)}
+K_{ij} \varphi_{0} \; ,
\eqno (1)
$$
where the stroke denotes three-dimensional covariant 
differentiation tangentially with respect to the intrinsic
Levi-Civita connection of the boundary, whilst $K_{ij}$ is
the extrinsic-curvature tensor of the boundary (we assume 
that $K_{ij}$ is nowhere vanishing). By virtue of (1),
the boundary conditions
$$
\Bigr[h_{ij}\Bigr]_{\partial M}=0
\eqno (2)
$$
are also obeyed by ${ }^{\varphi}h_{ij}$ if and only if
the whole ghost one-form obeys homogeneous Dirichlet
conditions:
$$
\Bigr[\varphi_{a} \Bigr]_{\partial M}=0 \; , \;
\forall a=0,1,2,3 \; .
\eqno (3)
$$
The problem now arises to impose boundary conditions on the
remaining set of metric perturbations, in such a way that
their invariance under infinitesimal diffeomorphisms is
again guaranteed by (3), since otherwise one would obtain
incompatible sets of boundary conditions on the ghost
one-form. For this purpose, one can point out that, if
$\Phi_{a}$ is any gauge-averaging functional which leads to
self-adjoint elliptic operators on metric perturbations,
one finds
$$
\delta \Phi_{a}(h) \equiv \Phi_{a}(h)
-\Phi_{a}({ }^{\varphi}h)={\cal F}_{a}^{\; \; b} \;
\varphi_{b} \; ,
\eqno (4)
$$
where ${\cal F}_{a}^{\; \; b}$ is an elliptic operator that 
acts linearly on the ghost one-form. Thus, if one imposes 
the boundary conditions
$$
\Bigr[\Phi_{a}(h)\Bigr]_{\partial M}=0 \; , \;
\forall a=0,1,2,3 \; ,
\eqno (5)
$$
their gauge invariance is guaranteed when (3) holds, by virtue
of (4). Hence one also has 
$\Bigr[\Phi_{a}({ }^{\varphi}h)\Bigr]_{\partial M}=0$,
$\forall a=0,1,2,3$.

If the axial (A) gauge-averaging functional is used:
$\Phi_{a}^{(A)}(h) \equiv n^{b}h_{ab}$, a considerable
simplification of the boundary conditions is obtained, since
all $h_{ab}$ perturbations are then set to zero at
$\partial M$. The resulting ghost operator takes the form [2]
$$
{\cal F}_{a}^{\; \; b}=2 \delta_{a}^{\; \; (b} \;
n^{c)} \; \nabla_{c} \; .
\eqno (6)
$$
This implies that the ghost operator does not have any
eigenfunctions at all. Indeed, given the eigenvalue equation
${\cal F} \varphi_{\lambda}=\lambda \varphi_{\lambda}$, its 
solution in the coordinates $\tau,{\hat x}$ ($\tau$ being a
radial coordinate, and $\hat x$ local coordinates on
$\partial M$) is [2]
$$
\varphi_{0_{\lambda}}(\tau,{\hat x})
=\exp \Bigr({1\over 2}\lambda \tau \Bigr)
f_{0_{\lambda}}({\hat x}) \; ,
\eqno (7)
$$
$$
\eqalignno{
\; & \varphi_{i_{\lambda}}(\tau,{\hat x})=
\exp(\lambda \tau) g_{ij}(\tau,{\hat x})
f_{\lambda}^{j}({\hat x}) \cr
&-\int_{0}^{\tau}dy \; \exp \biggr[\lambda
\Bigr(\tau-{1\over 2}y \Bigr)\biggr]
g_{ij}(\tau,{\hat x})g^{jk}(y,{\hat x})
{\widehat \nabla}_{k}f_{0_{\lambda}}({\hat x}) \; .
&(8)\cr}
$$
Now imposing the boundary conditions (3) one finds
$f_{0_{\lambda}}=f_{\lambda}^{i}=0$, and hence 
$\varphi_{\lambda}=0$ for any $\lambda$. Thus, ghost 
fields do not contribute at all to the transition 
amplitudes in our problem.

The remaining part of the analysis in the axial gauge
is as follows [2]. The spectrum of the operator on metric
perturbations can be obtained by studying the spectrum 
of the operator (here $\cstok{\ } \equiv g^{ab} 
\nabla_{a} \nabla_{b}$)
$$
\eqalignno{
\Delta^{ab,cd}&=-\Bigr(g^{a(c} \; g^{d)b}
-g^{ab}g^{cd}\Bigr)\cstok{\ }
-g^{cd}\nabla^{(a} \; \nabla^{b)} \cr
&-g^{ab}\nabla^{(c} \; \nabla^{d)}
+2 \nabla^{(a} \; g^{b)(c} \; \nabla^{d)} \; .
&(9)\cr}
$$
Covariant differentiation of the resulting eigenvalue 
equation, multiplication by $g^{ab}$ with summation over
repeated indices, and imposition of the axial gauge imply
that the only non-vanishing metric perturbations are the
three-dimensional transverse-traceless tensors. This
remains true if one performs a Gaussian average over the
axial-type functional. The integrability condition for the
eigenvalue equation contains then further contributions 
resulting from the extrinsic-curvature tensor, and selects
three-dimensional transverse-traceless perturbations,
provided that the unperturbed extrinsic-curvature tensor
is proportional to the three-metric of the boundary [2].
The full one-loop divergence is then given by 
$\zeta(0)=-{278\over 45}$, if one studies a portion of
flat Euclidean four-space bounded by a 
three-sphere [2]. This is a non-trivial
property, since a gauge has been found such that the
contributions of ghost and gauge modes vanish separately
in the presence of boundaries. This property is not shared
by other non-covariant gauges, e.g. the Coulomb gauge for
Euclidean Maxwell theory, where the ghost and gauge 
contributions cannot be made to vanish separately for
problems with boundary [2]. 

The axial gauge has been recently applied to the one-loop
semiclassical analysis of simple supergravity in the
presence of boundaries [1,3]. This analysis shows that the
effects of three-dimensional transverse-traceless perturbations
for gravitons and gravitinos do not cancel each other exactly,
and hence simple supergravity is one-loop finite just in those
particular backgrounds with boundary where pure gravity is
one-loop finite as well [4].
\vskip 0.3cm
\leftline {\bf Acknowledgments}
\vskip 0.3cm
\noindent
The work of I.G. Avramidi was supported by the Alexander von
Humboldt Foundation, by the Istituto Nazionale di Fisica
Nucleare and by the Deutsche Forschungsgemeinschaft. G. Esposito
is grateful to the European Union for financial support under
the Human Capital and Mobility Programme. A.Yu. Kamenshchik
was partially supported by the Russian Foundation for
Fundamental Researches through grant no 96-02-16220-a,
and by the Russian Research Project ``Cosmomicrophysics".
\vskip 0.3cm
\leftline {\bf References}
\vskip 0.3cm
\item {[1]}
G. Esposito, A.Yu. Kamenshchik and G. Pollifrone, Euclidean
quantum gravity on manifolds with boundary (submitted to
Kluwer Academic, 1996).
\item {[2]}
I.G. Avramidi, G. Esposito and A.Yu. Kamenshchik, Class.
Quantum Grav. 13 (1996) 2361.
\item {[3]}
G. Esposito and A.Yu. Kamenshchik, Phys. Rev. D
54 (1996) 3869.
\item {[4]}
G. Esposito, Local boundary conditions in quantum supergravity,
DSF preprint 96/14, HEP-TH 9608076 (to be published in
Phys. Lett. B).

\bye